\documentclass[twocolumn]{revtex4}
\usepackage{epsfig}
\usepackage{graphicx}
\usepackage[tbtags]{amsmath}

\begin{document}

\title{Proposal for Direct, Local Measurement of Entanglement for Pure Bipartite Systems of Arbitrary Dimension}
\author{Sang Min Lee and Hai-Woong Lee}
\address{Department of Physics, Korea Advanced Institute of Science and Technology, Daejeon 305-701, Korea}

\begin{abstract}
Based on the complementarity relation between entanglement of a composite system and the purity of a subsystem, we propose a simple method to measure the amount of entanglement. The method can be applied to a bipartite system in a pure state of any arbitrary dimension. It requires only single qudit rotations and straightforward probability measurements performed on one of the subsystems, and can thus be easily implemented experimentally using linear optical devices.
\end{abstract}

\maketitle

One of the key issues in quantum information science is how to detect and measure entanglement. Perhaps the most straightforward way of measuring entanglement is full tomographic reconstruction of the quantum state in question. This technique is, however, highly inefficient and time-consuming. In recent years several less demanding methods of detecting and measuring entanglement have been proposed [1-7] and some demonstrated experimentally [8-10]. These methods typically require measurements performed on the entire composite system and involve coincidence measurements and/or complex controlled operations. Some methods require measurements upon two copies of the quantum state.

On the other hand, it has been known that the amount of entanglement manifests itself in certain properties, such as the purity, of a subsystem [4,11,12]. Thus, the method of minimal and optimal tomography [13] on one of the subsystems can be used to measure entanglement [14]. Simpler ways of measuring the purity of a subsystem and consequently the entanglement exist [15,16]. These methods, however, are not without experimental difficulties. The method proposed by Ekert el al. [15] requires a controlled SWAP operation for which two identical copies of the state, whose purity is to be determined, need to be provided as target qudits. The method of D\"{u}rr [16] requires probability measurements with multiport beam splitters, in which probabilities need to be measured by varying phases associated with each port of the beam splitters over the entire range of the phase angles.

In this paper we propose a simple method of measuring bipartite entanglement. The quantity to be measured is the concurrence [17] or its generalized version known as the $I$ concurrence [11]. The system whose entanglement is to be measured needs to be in a pure state. The proposed method has several notable features. First, it requires measurements to be made only on one of the subsystems. Second, it requires only single qudit rotations and straightforward projection measurements. Third, it can be applied to systems of any arbitrary dimension. We also describe how our method can be experimentally implemented using linear optical devices.

The starting point of our analysis is the complementarity relation [11,12,18] between the $I$ concurrence $\mathcal{C}_{AB}$ of a composite quantum system $AB$ and the purity $\textrm{Tr}_{A}(\rho_{A} ^2)$ of a subsystem $A$ (or $B$), which reads, for the case when the composite system is in a pure state,
\begin{align}\label{e1}
\mathcal{C}^2_{AB}=2\left[1-\textrm{Tr}_A(\rho^2_A)\right]=2\left[1-\textrm{Tr}_B(\rho^2_B)\right]
\end{align}
The amount of entanglement can thus be determined if the purity $\textrm{Tr}(\rho^2)$ of one of the subsystems is measured. As the purity can be written as
\begin{align}\label{e2}
\textrm{Tr}(\rho^2)=\sum^d_{i=1}\rho^2_{ii}+2\sum_{(i,j)}\left|\rho_{ij}\right|^2
\end{align}
where $d$ is the dimension of the subsystem being measured, $d = d_A$ (or $d=d_B$) [From now on, we drop the subscript $A$ (or $B$). It is to be understood that all quantities refer to the subsystem being measured.], and $\sum_{(i,j)}$ means summation over all possible $\frac{d(d-1)}{2}$ pairs of the indices $i$ and $j$ with $i<j$, the determination of the $I$ concurrence requires knowledge on both diagonal and off-diagonal elements of the density matrix of the subsystem. The diagonal element $\rho_{ii}$ is identified as the probability  $P_i$ of finding the subsystem in state $|i\rangle$ and can thus be determined directly from the projection measurement,
\begin{align}\label{e3}
P_i=\rho_{ii}; \ \ \ \ \ \ \ \ \  i=1, 2, \cdots, d.
\end{align}

In order to determine the off-diagonal elements, we first need to perform a $90^\circ$ rotation of the state of the subsystem about $x$ axis in the clockwise direction in the three-dimensional Bloch sphere representation of the two-dimensional $|k\rangle-|l\rangle$ space ($k < l; \  k,l=1, 2, \cdots,d$). The density matrix of the rotated system is given by $\rho'=U(k,l)\rho U(k,l)^\dagger$, where $U(k,l)$ is a $d \times d$ rotation matrix with the elements given by
\begin{gather}\label{e4}
U_{kk}=U_{ll}=\frac{1}{\sqrt{2}}, \ \ \ \ U_{kl}=U_{lk}=\frac{i}{\sqrt{2}}, \notag \\
U_{nn}=1 \ (n\neq k,l), \ \ \ \ U_{nm}=0 \ (n\neq m ; \ nm \neq kl, lk).
\end{gather}
The projection measurement performed on the rotated system yields the probabilities
\begin{align}\label{e5}
&P_k'=\rho'_{kk}=\frac{1}{2}\left[\rho_{kk}+\rho_{ll}-i(\rho_{kl}-\rho_{lk}) \right], \notag\\
&P_l'=\rho'_{ll}=\frac{1}{2}\left[\rho_{kk}+\rho_{ll}+i(\rho_{kl}-\rho_{lk}) \right], \notag\\
&P_n'=\rho'_{nn}=\rho_{nn}; \ \ \ \ n\neq k,l.
\end{align}
Since we know the diagonal elements $\rho_{kk}$ and $\rho_{ll}$ from the straight projection measurement [Eq.(3)], we can determine the imaginary part of the off-diagonal element $\rho_{kl}$, $\textrm{Im}(\rho_{kl})=\frac{-i}{2}(\rho_{kl}-\rho_{lk})$, by measuring the probabilities $P'_k$ and $P'_l$. This can be understood by noting that physically the $90^\circ$ rotation about $x$ axis projects the $y$ component of the Bloch vector, which is equivalent to the imaginary part of the off-diagonal element, onto $z$ axis. Performing the rotation $U$ and the projection measurement on each of $\frac{d(d-1)}{2}$ pairs of the states $|k\rangle$ and $|l\rangle$, we can determine the imaginary part of all $\frac{d(d-1)}{2}$ off-diagonal elements $\rho_{ij}$.

Having determined the imaginary part of the off-diagonal elements, we next need to determine the real part. That can be accomplished by a $90^\circ$ rotation about $y$ axis in the Bloch space representing the two states $|k\rangle$ and $|l\rangle$. The rotation brings the state to $\rho''=V(k,l)\rho V(k,l)^{\dagger}$, where
\begin{gather}\label{e6}
V_{kk}=V_{ll}=\frac{1}{\sqrt{2}}, \ \ \ \ V_{kl}=\frac{1}{\sqrt{2}}, \ \ \ \ V_{lk}=-\frac{1}{\sqrt{2}}, \notag \\
V_{nn}=1 \ (n\neq k,l), \ \ \ \ V_{nm}=0 \ (n\neq m ; \ nm \neq kl, lk).
\end{gather}
The projection measurement on the rotated system yields now the probabilities
\begin{align}\label{e7}
&P''_k=\rho''_{kk}=\frac{1}{2}\left[\rho_{kk}+\rho_{ll}+(\rho_{kl}+\rho_{lk}) \right], \notag\\
&P''_l=\rho''_{ll}=\frac{1}{2}\left[\rho_{kk}+\rho_{ll}-(\rho_{kl}+\rho_{lk}) \right], \notag\\
&P''_n=\rho''_{nn}=\rho_{nn}; \ \ \ \ n\neq k,l.
\end{align}
The real part of the off-diagonal element $\rho_{kl}$, $\textrm{Re}(\rho_{kl})=\frac{1}{2}(\rho_{kl}+\rho_{lk})$, can thus be determined from the measurement of the probabilities $P''_k$ and $P''_l$. The $90^\circ$ rotation about $y$ axis projects the $x$ component of the Bloch vector onto $z$ axis, allowing the real part of $\rho_{kl}$ to be determined. As before we need to perform the rotation $V$ and the projection measurement on each of $\frac{d(d-1)}{2}$ pairs of the states $|k\rangle$ and $|l\rangle$ to determine the real part of all $\frac{d(d-1)}{2}$ off-diagonal elements $\rho_{ij}$.

Eq.(2) indicates that the determination of the purity requires the sum of the squares of all diagonal and off-diagonal elements, corresponding to the squares of the length of the Bloch vector projected onto $z$ axis and onto $xy$ plane, respectively. The sum of the squares of the diagonal elements can be obtained by squaring the probabilities $P_i$ obtained from the straight projection measurement and summing them all,
\begin{align}\label{e8}
T=\sum^{d}_{i=1}P^2_i=\sum^{d}_{i=1}\rho^2_{ii}.
\end{align}
Information on the sum of the squares of the off-diagonal elements is contained in the quantity $T'+T''$ where
\begin{gather}\label{e9}
T'=\sum_{(k,l)}\sum^d_{i=1}P'^2_i, \notag \\
T''=\sum_{(k,l)}\sum^d_{i=1}P''^2_i.
\end{gather}
Here $\sum_{(k,l)}$ means summation over all possible  $\frac{d(d-1)}{2}$ pairs of the states $|k\rangle$ and $|l\rangle$ with $k<l$. A straightforward calculation yields
\begin{align}\label{e10}
T'+T''=(d^2-2d)T+2\sum_{(i,j)}\left|\rho_{ij}\right|^2+1.
\end{align}
Combining Eqs. (2), (8) and (9), we obtain for the purity
\begin{align}\label{e11}
\textrm{Tr}(\rho^2)=T'+T''-(d^2-2d-1)T-1.
\end{align}
The $I$ concurrence is then given according to Eq. (1) by
\begin{align}\label{e12}
\mathcal{C}_{AB}=\sqrt{4+2(d^2-2d-1)T-2(T'+T'')}.
\end{align}

We now describe an experimental scheme that actually performs the measurement. We adopt a multipath representation of a qudit. In a multipath system any arbitrary $d \times d$ unitary operator can be constructed using an arrangement of beam splitters, phase shifters and mirrors [19]. In particular, the rotation operator $U(k,l)$ of Eq. (4) can be realized by a 50/50 beam splitter situated at the intersection of the $k$th and $l$th paths as shown in Fig.1. The state transformation performed by the beam splitter is represented by the relation
\begin{gather}\label{e}
\begin{pmatrix} k_{out} \\ l_{out}\end{pmatrix}
=\frac{1}{\sqrt{2}} \begin{pmatrix} 1 & i \\ i & 1 \end{pmatrix}
\begin{pmatrix} k_{in} \\ l_{in}\end{pmatrix},\\
n_{out}=n_{in}; \ \ \ \ n\neq k,l. \notag
\end{gather}
On the other hand, the rotation operator $V(k,l)$ of Eq.(6) represented by the relation
\begin{gather}\label{e}
\begin{pmatrix} k_{out} \\ l_{out}\end{pmatrix}
=\frac{1}{\sqrt{2}} \begin{pmatrix} 1 & 1 \\ -1 & 1 \end{pmatrix}
\begin{pmatrix} k_{in} \\ l_{in}\end{pmatrix},\\
n_{out}=n_{in}; \ \ \ \ n\neq k,l. \notag
\end{gather}
can be realized by a beam splitter at the intersection of the $k$th and $l$th paths and two phase shifters that shift phase by $-\frac{\pi}{2}$ and $\frac{\pi}{2}$, respectively, situated at the input port and the output port, respectively of the $l$th path, as shown in Fig.1.

\begin{figure}[h]
\centerline{\includegraphics[scale=0.6]{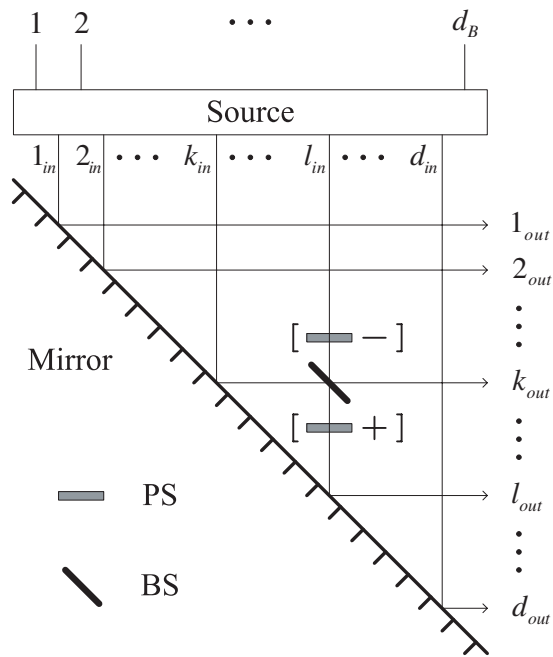}}
\caption{Rotation $U(k,l)$[$V(k,l)$] accomplished by a beam splitter (BS) placed at the intersection of the $k$th and $l$th paths [and a $-\frac{\pi}{2}$ phase shifter ($-$) and a $\frac{\pi}{2}$ phase shifter ($+$) at the input port and the output port, respectively, of the $l$th path].}
\label{f1}
\end{figure}

Our experiment then proceeds as follows. We assume that a source emits a large number of pairs of path-entangled particles $AB$ of dimension $d_A \times d_B = d \times d_B$ (see Fig.1). The $I$ concurrence of the composite system $AB$ is to be determined by performing local measurements on one of the subsystems, e.g., the subsystem $A$. The subsystem being measured has the dimension of $d_A=d$.\\
(1) The projection measurement is performed directly on the subsystem $A$ to measure the probability $P_i$ for the subsystem $A$ to be in the $i$th path. The measurement can be achieved with the arrangement shown in Fig.1 with the beam splitter and the phase shifters taken out. Calculate $T=\sum^{d}_{i=1}P^2_i$.\\
(2) The projection measurement is performed with a beam splitter placed at the intersection of the $k$th and $l$th paths to determine the probability $P'_i$. The measurement can be achieved with the arrangement shown in Fig.1 with the beam splitter kept installed but the phase shifters taken out. Repeat the projection measurement with the beam splitter shifted to the intersection of all other possible pairs of paths. Calculate $T'=\sum_{(k,l)}\sum^d_{i=1}P'^2_i$\\
(3) The projection measurement is performed with a beam splitter and two phase shifters of $-\frac{\pi}{2}$ and $\frac{\pi}{2}$, respectively, placed at the intersection of the $k$th and $l$th paths to determine the probability $P''_i$. The measurement can be achieved with the arrangement shown in Fig.1 with the beam splitter and the phase shifters kept installed. Repeat the projection measurement with the beam splitter and the two phase shifters shifted to the intersection of all other possible pairs of paths. Calculate $T''=\sum_{(k,l)}\sum^d_{i=1}P''^2_i$.\\
(4) Determine the $I$ concurrence using Eq.(12).

The number of experimental setups required to complete the above experiment is 1 for step (1) and $\frac{d(d+1)}{2}$ each for steps (2) and (3), totaling $d^2-d+1$. This number can considerably be reduced by placing beam splitters at every possible intersection simultaneously for steps (2) and (3), which increases the number of the real and imaginary parts, respectively, of the off-diagonal elements that can be determined from one experimental setup to $\frac{d}{2}$ for even $d$ and $\frac{d-1}{2}$ for odd $d$. Each experimental setup for steps (2) and (3), respectively, then produces state $\rho'$ and $\rho''$ given by $\rho'=U(k_{N},l_{N})\cdots U(k_{1},l_{1})\rho U(k_{1},l_{1})^\dagger\cdots U(k_{N},l_{N})^\dagger$ and $\rho''=V(k_{N},l_{N})\cdots V(k_{1},l_{1})\rho V(k_{1},l_{1})^\dagger\cdots V(k_{N},l_{N})^\dagger$, where $N=\frac{d}{2}$ for even $d$ and $N=\frac{d-1}{2}$ for odd $d$. This way the total number of experimental setups required reduces to $2d-1$ for even $d$ and $2d+1$ for odd $d$. A straightforward calculation shows that the $I$ concurrence is then given by
\begin{gather}
\mathcal{C}_{AB}=\left\{
\begin{array}{lc}
\sqrt{4 + 2 (d-3)T - 2 (T'+T'') } & \textrm{for even} \ d\\
\sqrt{4 + 2 (d-1) T - 2 (T'+T'')} & \textrm{for odd} \ d
\end{array}
\right.
\end{gather}

We illustrate our scheme for the simple case of a two-qubit system in state $|\psi \rangle_{AB}=\cos \theta |01\rangle_{AB}+\sin \theta |10\rangle_{AB}$. Let us choose the subsystem $A$ to be the subsystem to be measured. Since $\rho\equiv\rho_A=\cos^2 \theta |0\rangle\langle0| + \sin^2 \theta |1\rangle\langle1| = \begin{pmatrix}\cos^2 \theta & 0 \\ 0 & \sin^2 \theta \end{pmatrix}$, the direct projection measurement should yield $P_1=\cos^2 \theta$ and $P_2=\sin^2 \theta$, and thus $T=\cos^4 \theta + \sin^4 \theta$. (Here the subscripts 1 and 2 refer to states $|0\rangle$ and $|1\rangle$, respectively.) It can be easily seen that $\rho'=U(1,2)\rho U(1,2)^\dagger = \frac{1}{2} \begin{pmatrix} 1 & -i \cos 2\theta \\ i \cos 2\theta & 1 \end{pmatrix}$ and $\rho''=V(1,2)\rho V(1,2)^\dagger = \frac{1}{2} \begin{pmatrix} 1 & - \cos 2\theta \\ - \cos 2\theta & 1 \end{pmatrix}$. Thus, the projection measurements of steps (2) and (3) should yield $P'_1=\frac{1}{2}$, $P'_2=\frac{1}{2}$ and $P''_1=\frac{1}{2}$, $P''_2=\frac{1}{2}$, and therefore $T'=T''=\frac{1}{2}$. The $I$ concurrence for this case is the usual concurrence and is given according to Eq.(12) [or Eq. (15)] by $\mathcal{C}_{AB}=|\sin 2\theta|$. If the qubit is a photon polarization qubit, the direct projection measurement corresponds to a polarization measurement in the horizontal/vertical basis, while the projection measurement following the rotation $U$ and $V$, respectively, corresponds to a polarization measurement in the 45$^\circ$/-45$^\circ$ basis and in the right-circular/left-circular basis.

For the general two-qubit state $|\psi \rangle_{AB}=a|00\rangle_{AB}+b|01\rangle_{AB}+c|10\rangle_{AB}+d|11\rangle_{AB}$, we obtain, through a straightforward calculation, $P_1=|a|^2+|b|^2$, $P_2=|c|^2+|d|^2$, $P'_1=\frac{1}{2}+\textrm{Im}(ac^\ast+bd^\ast)$, $P'_2=\frac{1}{2}-\textrm{Im}(ac^\ast+bd^\ast)$, $P''_1=\frac{1}{2}+\textrm{Re}(ac^\ast+bd^\ast)$ and $P''_2=\frac{1}{2}-\textrm{Re}(ac^\ast+bd^\ast)$. Eq.(12) then immediately yields $\mathcal{C}_{AB}=2|ad-bc|$.

In summary we propose a simple experimental scheme that allows one to determine the amount of entanglement of a bipartite composite system in a pure state. The scheme requires local probability measurements only on one of the subsystems. Furthermore, the scheme requires only single qudit rotations with linear optical devices and straightforward projection measurements. No complex controlled operations are needed. The method can be applied to systems of any arbitrary dimension $d_A \times d_B$. The number of required experimental setups scales linearly with the dimension of the subsystem being measured. The scheme can be effective especially when one of the subsystems has a small dimension. If the composite system is in a mixed state, Eq. (1) takes the form of inequality and our proposed method can only determine the upper bound of the $I$ concurrence.

This research was supported by a grant from the Korea Research Institute for Standards and Science (KRISS).


\begin{references}

\bibitem{1} Pawe{\l} Horodecki and Artur Ekert, Phys. Rev. Lett. \textbf{89}, 127902 (2002); Pawe{\l} Horodecki, Phys. Rev. Lett. \textbf{90}, 167901 (2003).

\bibitem{2} O. G\"{u}hne, P. Hyllus, D. Bru{\ss}, A. Ekert, M. Lewenstein, C. Macchiavello, and A. Sanpera, Phys. Rev. A \textbf{66}, 062305 (2002).

\bibitem{3} Jian-Ming Cai, Zheng-Wei Zhou, and Guang-Can Guo, Phys. Rev. A \textbf{73}, 024301 (2006).

\bibitem{4} Christian Kothe and Gunnar Bj\"{o}rk, Phys. Rev. A \textbf{75}, 012336 (2007).

\bibitem{5} Isabel Sainz Abascal and Gunnar Bj\"{o}rk, Phys. Rev. A \textbf{75}, 062317 (2007).

\bibitem{6} Alexander Klyachko, Bar{\i}\c{s} \"{O}ztop, and Alexander S. Shumovsky, Appl. Phys. Lett. \textbf{88}, 124102 (2006); Alexander Klyachko, Bar{\i}\c{s} \"{O}ztop, and Alexander S. Shumovsky, Phys. Rev. A \textbf{75}, 032315 (2007).

\bibitem{7} Sang Min Lee, Se-Wan Ji, Hai-Woong Lee, and M. Suhail Zubairy, Phys. Rev. A \textbf{77}, 040301(R) (2008).

\bibitem{8} M. Barbieri, F. De Martini, G. Di Nepi, P. Mataloni, G. M. D¡¯Ariano, and C. Macchiavello, Phys. Rev. Lett. \textbf{91}, 227901 (2003).

\bibitem{9} S. P. Walborn, P. H. Souto Ribeiro, L. Davidovich, F. Mintert, and A. Buchleitner, Nature(London) \textbf{440}, 1022 (2006).

\bibitem{10} Zhi-Wei Wang, Yun-Feng Huang, Xi-Feng Ren, Yong-Sheng Zhang, and Guang-Can Guo, Europhys. Lett. \textbf{78}, 40002 (2007).

\bibitem{11} Pranaw Rungta, V. Bu\v{z}ek, Carlton M. Caves, M. Hillery, and G. J. Milburn, Phys. Rev. A \textbf{64}, 042315 (2001).

\bibitem{12} Matthias Jakob and J\'{a}nos A. Bergou, Phys. Rev. A \textbf{76}, 052107 (2007).

\bibitem{13} Jaroslav \v{R}eh\'{a}\v{c}ek, Berthold-Georg Englert, and Dagomir Kaszlikowski, Phys. Rev. A \textbf{70}, 052321 (2004).

\bibitem{14} A. Salles, F. de Melo, J. C. Retamal, R. L. de Matos Filho, and N. Zagury, Phys. Rev. A \textbf{74}, 060303(R) (2006).

\bibitem{15} Artur K. Ekert, Carolina Moura Alves, and Daniel K. L. Oi, Phys. Rev. Lett. \textbf{88}, 217901 (2002).

\bibitem{16} Stephan D\"{u}rr, Phys. Rev. A \textbf{64}, 042113 (2001).

\bibitem{17} William K. Wootters, Phys. Rev. Lett. \textbf{80}, 2245 (1998).

\bibitem{18} Gregg Jaeger, Michael A. Horne, and Abner Shimony, Phys. Rev. A \textbf{48}, 1023 (1993); Gregg Jaeger, Abner Shimony, and Lev Vaidman, Phys. Rev. A \textbf{51}, 54 (1995).

\bibitem{19} Michael Reck, Anton Zeilinger, Herbert J. Bernstein, and Philip Bertani, Phys. Rev. Lett. \textbf{73}, 58 (1994).

\end{references}
\end{document}